\documentclass[11pt,fullpage]{article}

\newcommand{\ol}{\setlength{\itemsep}{0pt.}\begin{enumerate}}
\newcommand{\eol}{\end{enumerate}\setlength{\itemsep}{-\parsep}}
\newcommand{\ignore}[1]{}
\title{\bf A proof of the log-concavity conjecture related to the computation of the ergodic capacity of
 {\bf MIMO} channels}
 
\author{Leonid Gurvits \thanks{%
{\tt gurvits@lanl.gov}. Los Alamos National Laboratory, 
Los Alamos, NM. } 
}

\begin{document}



\maketitle

\begin{abstract}
An upper bound on the ergodic capacity of {\bf MIMO} channels was introduced recently in \cite{gao}. This upper
bound amounts to the maximization on the simplex of some multilinear polynomial $p(\lambda_1,...,\lambda_n)$ with non-negative coefficients.
In general, such maximizations problems are {\bf NP-HARD}. But if say, the functional $\log(p)$ is concave
on the simplex and can be efficiently evaluated, then the maximization
can also be done efficiently. Such log-concavity was conjectured in \cite{gao}. 
We give in this paper self-contained proof of the conjecture, based on the theory of
{\bf H-Stable} polynomials. 

\end{abstract} 

 
\newtheorem{THEOREM}{Theorem}[section]
\newenvironment{theorem}{\begin{THEOREM} \hspace{-.85em} {\bf :} 
}%
                        {\end{THEOREM}}
\newtheorem{LEMMA}[THEOREM]{Lemma}
\newenvironment{lemma}{\begin{LEMMA} \hspace{-.85em} {\bf :} }%
                      {\end{LEMMA}}
\newtheorem{COROLLARY}[THEOREM]{Corollary}
\newenvironment{corollary}{\begin{COROLLARY} \hspace{-.85em} {\bf 
:} }%
                          {\end{COROLLARY}}
\newtheorem{PROPOSITION}[THEOREM]{Proposition}
\newenvironment{proposition}{\begin{PROPOSITION} \hspace{-.85em} 
{\bf :} }%
                            {\end{PROPOSITION}}
\newtheorem{DEFINITION}[THEOREM]{Definition}
\newenvironment{definition}{\begin{DEFINITION} \hspace{-.85em} {\bf 
:} \rm}%
                            {\end{DEFINITION}}
\newtheorem{EXAMPLE}[THEOREM]{Example}
\newenvironment{example}{\begin{EXAMPLE} \hspace{-.85em} {\bf :} 
\rm}%
                            {\end{EXAMPLE}}
\newtheorem{CONJECTURE}[THEOREM]{Conjecture}
\newenvironment{conjecture}{\begin{CONJECTURE} \hspace{-.85em} 
{\bf :} \rm}%
                            {\end{CONJECTURE}}
\newtheorem{PROBLEM}[THEOREM]{Problem}
\newenvironment{problem}{\begin{PROBLEM} \hspace{-.85em} {\bf :} 
\rm}%
                            {\end{PROBLEM}}
\newtheorem{QUESTION}[THEOREM]{Question}
\newenvironment{question}{\begin{QUESTION} \hspace{-.85em} {\bf :} 
\rm}%
                            {\end{QUESTION}}
\newtheorem{REMARK}[THEOREM]{Remark}
\newenvironment{remark}{\begin{REMARK} \hspace{-.85em} {\bf :} 
\rm}%
                            {\end{REMARK}}
\newtheorem{FACT}[THEOREM]{Fact}
\newenvironment{fact}{\begin{FACT} \hspace{-.85em} {\bf :} 
\rm}%
		            {\end{FACT}}

 
\newcommand{\thm}{\begin{theorem}}
\newcommand{\lem}{\begin{lemma}}
\newcommand{\pro}{\begin{proposition}}
\newcommand{\dfn}{\begin{definition}}
\newcommand{\rem}{\begin{remark}}
\newcommand{\xam}{\begin{example}}
\newcommand{\cnj}{\begin{conjecture}}
\newcommand{\prb}{\begin{problem}}
\newcommand{\que}{\begin{question}}
\newcommand{\cor}{\begin{corollary}}
\newcommand{\fac}{\begin{fact}}

\newcommand{\prf}{\noindent{\bf Proof:} }
\newcommand{\ethm}{\end{theorem}}
\newcommand{\elem}{\end{lemma}}
\newcommand{\epro}{\end{proposition}}
\newcommand{\edfn}{\bbox\end{definition}}
\newcommand{\erem}{\bbox\end{remark}}
\newcommand{\exam}{\bbox\end{example}}
\newcommand{\ecnj}{\bbox\end{conjecture}}
\newcommand{\eprb}{\bbox\end{problem}}
\newcommand{\eque}{\bbox\end{question}}
\newcommand{\ecor}{\end{corollary}}
\newcommand{\efac}{\end{fact}}
\newcommand{\eprf}{\bbox}
\newcommand{\beqn}{\begin{equation}}
\newcommand{\eeqn}{\end{equation}}
\newcommand{\wbox}{\mbox{$\sqcap$\llap{$\sqcup$}}}
\newcommand{\bbox}{\vrule height7pt width4pt depth1pt}
\newcommand{\qed}{\bbox}

\newcommand{\rarrow}{\rightarrow}
\newcommand{\larrow}{\leftarrow}
\newcommand{\grad}{\bigtriangledown}

\overfullrule=0pt
\def\setof#1{\lbrace #1 \rbrace}
\section{The conjecture}
Let $B$ be $M \times M$ matrix. Recall the definition of the {\bf permanent} :
$$
Per(B) = \sum_{\sigma \in S_M} \prod_{1 \leq i \leq M} A(i, \sigma(i)).
$$
The following Conjecture was posed in \cite{gao}.
\cnj \label{main-con}
Let $A$ be $M \times N, M < N$ matrix with non-negative entries. We denote as $A_{S}$ a submatrix
$$
A_{S} = \{A(i,j): 1 \leq i \leq m; j \in S \subset \{1,...,N\}.
$$
Define the following multi-linear polynomial with non-negative coefficients
\beqn \label{start}
F_{A}(\lambda_1,...,\lambda_N) = \sum_{|S| = M, S \subset \{1,...,N\} } Per(A_{S}) \prod_{j \in S} \lambda_j.
\eeqn
Then the functional $\log(F_{A})$ is concave on $R_{+}^{N} = \{(\lambda_1,...,\lambda_N) : \lambda_j \geq 0, 1 \leq j \leq N\}$.
\ecnj

We present in this paper a proof of Conjecture(\ref{main-con}). Actually we prove that the polynomial $F_{A}$ is either zero
or {\bf H-Stable}.

\section{ {\bf H-Stable} polynomials }
{\it To make this note self-contained, we present in this section proofs of a few necessary results. The reader may consult \cite{ejc}
and \cite{schr2} for the further reading and references.}

We denote as $Hom_{+}(m,n)$ a convex closed cone of homogeneous polynomials with non-negative coefficients of degree $n$ in $m$ variables
and with non-negative coefficients; as $R_{+}^m$ a convex closed cone of non-negative vectors in $R^m$ and
as $R_{++}^m$ a convex open cone of positive vectors in $R^m$.
\dfn
A homogeneous polynomial $p \in Hom_{+}(m,n)$ is called {\bf H-Stable} if
$$
|p(z_1,...,z_m)| > 0 ; Re(z_i) > 0, 1 \leq i \leq m;
$$
is called {\bf H-SStable} if $|p(z_1,...,z_m)| > 0$ provided that\\
$Re(z_i) \geq 0, 1 \leq i \leq m$ and $ 0 < \sum_{1 \leq m \leq m} Re(z_i)$.\\
\edfn
\xam \label{first}
Consider a bivariate homogeneous polynomial $p \in Hom_{+}(2,n)$, $p(z_1,z_2) = (z_2)^n P(\frac{z_1}{z_2})$,
where $P$ is some univariate polynomial. Then $p$ is {\bf H-Stable} iff the roots
of  $P$ are non-positive real numbers. This assertion is just a rephrasing of the next  set equality:\\
$$
{\bf C} - \{\frac{z_1}{z_2} : Re(z_1),Re(z_2) > 0\} = \{x \in R: x \leq 0\}.
$$
In other words 
$$
P(t) = a \prod_{1 \leq i \leq k \leq n} (t + a_i): a_i \geq 0, 1 \leq i \leq k; a > 0.
$$
Which gives the following expression for the bivariate homogeneous polynomial $p$:
$$
p(z_1,z_2) = a z_2^{n-k}\prod_{1 \leq i \leq k \leq n} (z_1 + a_i z_2)
$$
\exam
\fac \label{l-c}
Let $p \in Hom_{+}(m,n)$ be {\bf H-Stable}. Then $\log(p)$ is concave on $R_{+}^m$.
\efac
\prf
Consider two vectors $X,Y \in R_{+}^m$ such that their sum $X + Y \in R_{+}^m$ has all positive coordinates.
It is sufficient to prove that the bivariate homogeneous polynomial $q \in Hom_{+}(2,n)$
$$
q(t,s) = p(tX + sY),
$$
is log-concave on $R_{+}^2$. Clearly, the polynomial $q$ is {\bf H-Stable}. Therefore, using Example(\ref{first}),
we get that
$$
\log(q(t,s)) = \log(a) + (n-k) \log(s) + \sum_{1 \leq i \leq k \leq n} \log(t + a_i s):a_i \geq 0, 1 \leq i \leq k; a > 0.
$$
The log-concavity of $q$ follows now from the concavity of the logarithm on $[0, \infty)$.
\eprf
\rem
Since the polynomial $p$ is homogeneous of degree $n$ hence, by the standard argument, the function $ p^{\frac{1}{n}}$ is concave
on $R_{+}^m$ as well.
\erem
\fac
Let $p \in Hom_{+}(m,n)$ be {\bf H-Stable} and $x_i \geq 0, 1 \leq i \leq m$ then the following inequality holds
\beqn \label{max-prin}
|p(x_1 + i y_1,...,x_m + i y_m)| \geq p(x_1,...,x_m)
\eeqn
\efac
\prf
Consider without loss of generality the positive case $x_i > 0, 1 \leq i \leq m$.
Then there exists a positive real number $\mu > 0$ such that $y_i + \mu x_i > 0, 1 \leq i \leq m$. It follows from Example(\ref{first})
that for all complex numbers $z \in C$
$$
p(z x_1 + (y_1 + \mu x_1),..., x_m + z(y_m + \mu x_m) = p(x_1,...,x_m) \prod_{1 \leq i \leq n} (z + a_i); a_i > 0, 1 \leq i \leq m.
$$ 
Thus
$$
p(z x_1 + y_1,...,z x_m + y_m) =  p(x_1,...,x_m) \prod_{1 \leq i \leq n} (z + a_i - \mu)
$$
We get, using the homogeniuty of the polynomial $p$, that
$$
p(x_1 + i y_1,...,x_m + i y_m) = p(x_1,...,x_m) \prod_{1 \leq j \leq n} (1 + i (a_j - \mu)).
$$
As $|\prod_{1 \leq j \leq n} (1 + i (a_j - \mu))| \geq 1$ this proves that the inequality (\ref{max-prin}) holds.
\eprf
\cor
A nonzero polynomial $p \in Hom_{+}(m,n)$ is {\bf H-Stable} if and only the inequlity (\ref{max-prin}) holds.
\ecor

\cor \label{limit}
Let $p_i \in Hom_{+}(m,n)$ be a sequence of {\bf H-Stable} polynomials and $p = \lim_{i \rightarrow \infty} p_i$.
Then $p$ is either zero or {\bf H-Stable}.
\ecor
{\it Some readers might recognize Corollary (\ref{limit}) as a particular case of A. Hurwitz's theorem on limits
of sequences of nowhere zero analytical functions. Our proof below is elementary.}\\ 

\prf
Suppose that $p$ is not zero. Since $p \in Hom_{+}(m,n)$ hence $p(x_1,\dots,x_m) > 0$ if $x_j > 0: 1 \leq j \leq m$.
As the polynomials $p_i$ are {\bf H-Stable} therefore $|p_{i}(Z)| \geq |p_{i} \left(Re(Z) \right)|: Re(Z) \in R_{++}^{m}$.
Taking the limits we get that $|p(Z)| \geq |p \left(Re(Z) \right)| > 0: Re(Z) \in R_{++}^{m}$, which means
that $p$ is {\bf H-Stable}. 
\eprf

We need the following simple yet crucial result.
\pro \label{dif}
Let $p \in Hom_{+}(m,n)$ be {\bf H-Stable}. Then the polynomial $p_{(1)} \in Hom_{+}(m-1,n-1)$,
$$
p_{(1)}(x_2,...,x_m) =: \frac{\partial}{\partial x_{1}} p(0,x_2,...,x_m),
$$
is either zero or {\bf H-Stable}.
\epro
\prf 
Fix complex numbers $z_i, 2 \leq i \leq m$ and define the following univariate polynomial
$$
R(t) = p(t,z_2,...,z_m).
$$
It follows that $R^{\prime}(0) = p_{(1)}(z_2,...,z_m)$. We consider two cases.\\
First case: the polynomial $p \in Hom_{+}(m,n)$ is {\bf H-SStable}. In this case the polynomial $p_{(1)} \in Hom_{+}(m-1,n-1)$ is {\bf H-SStable} as well.
Indeed, in this case if the real parts $Re(z_i) \geq 0, 2 \leq i \leq m$ and $\sum_{2 \leq i \leq m} RE(z_i) > 0$ then all
the roots $v_1,...,v_{n-1}$ of the univariate polynomial $R$ have strictly negative real parts:
$$
R(t) = h \prod_{2 \leq i \leq n-1} (t - v_i), 0 \neq h \in C.
$$
Therefore
$$
p_{(1)}(z_2,...,z_m) = R^{\prime}(0) = h (-1)^{n-2} (\prod_{2 \leq i \leq n-1} v_i) (\sum_{2 \leq i \leq n-1} (v_i)^{-1}) \neq 0
$$
as the real part 
$$
Re(\sum_{2 \leq i \leq n-1} (v_i)^{-1}) = \sum_{2 \leq i \leq n-1} \frac{Re(v_i)}{|v_i|^2} > 0.
$$
Second case: the polynomial $p \in Hom_{+}(m,n)$ is {\bf H-Stable} but not {\bf H-SStable}. We need to approximate $p$
by a sequence of {\bf H-SStable} polynomials. Here is one natural approach: let $A$ be any $m \times m$ matrix with
positive entries. Define the following polynomials:
$$
p_{I + \epsilon A}(Z) =: p\left( (I + \epsilon A) Z \right), Z \in C^m.
$$
Clearly, the for all $\epsilon > 0$ the polynomials $p_{I + \epsilon A} \in Hom_{+}(m,n)$ and are {\bf H-SStable}.\\
It follows that polynomials $\frac{\partial}{\partial x_{1}} p_{I + \epsilon A}(0,x_2,...,x_m)$ are {\bf H-SStable} as well.
Note that
$$
\lim_{\epsilon \rightarrow 0} \frac{\partial}{\partial x_{1}} p_{I + \epsilon A}(0,z_2,...,z_m) = p_{(1)}(z_2,...,z_m).
$$
Using Corollary(\ref{limit}) we get that the polynomial $ p_{(1)}$ is either {\bf H-Stable} or zero.
\eprf

\section{Proof of the conjecture}
\prf
We will need a few auxillary polynomials:
\beqn
P(x_1,...,x_M; \lambda_1,...,\lambda_N) = \prod_{1 \leq j \leq N} (\lambda_j + \sum_{1 \leq i \leq m} A(i,j) x_i).
\eeqn 
Clearly, the polynomial $P \in Hom_{+}(M+N, N)$ is  {\bf H-Stable} if the entries of the matrix $A$ are non-negative. Applying Proposition(\ref{dif})
inductively, we get that the following
polynomial
\beqn \label{pa-1}
R(\lambda_1,...,\lambda_N) =\frac{\partial^{m}}{\partial x_{1}...\partial x_{m}} P(X=0;\lambda_1,...,\lambda_N)
\eeqn
is either zero or {\bf H-Stable} as well. It is easy to see that
\beqn
R(\lambda_1,...,\lambda_N) = \sum_{|S| = M, S \subset \{1,...,N\} } Per(A_{S}) \prod_{j \in \bar{S}} \lambda_j,
\eeqn \label{pa-2}
where $\bar{S} = \{1,...,N\} - S$ is the compliment of the set $S$.\\
Now everything is ready for the punch line: the {\bf multilinear homogeneous polynomial}, defined in (\ref{start}), 
\beqn \label{pa-3}
F_{A}(\lambda_1,...,\lambda_N) = (\prod_{1 \leq i \leq N}\lambda_i) R((\lambda_1)^{-1},...,(\lambda_N)^{-1}).
\eeqn
Recall that the real part $Re(z^{-1}) = \frac{Re(z)}{|z|^2}$ for all non-zero complex numbers $z \in C$.
Therefore, if the real parts $Re(\lambda_i) > 0, 1 \leq i \leq n$ then the same is true for the inverses:
$$
Re((\lambda_i)^{-1}) > 0, 1 \leq i \leq n.
$$
This proves that the polynomial $F_{A}$ is either zero or {\bf H-Stable}. The log-concavity follows from Fact(\ref{l-c}).  
\eprf

\section{Conclusion}
The reader should not be deceived by the simplicity of our proof: very similar arguments are behind the breakthrough results in \cite{ejc}, \cite{fried},
\cite{mixvol}. The reader is advised to read very nice exposition in  \cite{schr2}.\\
Conjecture (\ref{main-con}) is actually a very profound question. Had it been asked and properly answered in 1960-70s, then the theory of permanents
(and of related things like mixed discriminants and mixed volumes \cite{mixvol}) could have been very different now.\\
Though the ``permanental'' part in \cite{gao} is fairly standard(the authors essentially rediscovered
so called Godsil-Gutman Formula \cite{God-Gut}) it is quite amazing how naturally the permanent enters the story. Switching the expectation
and the logarithm can be eventful indeed.\\
The log-concavity comes up really handily in the optimizational context of \cite{gao}. The thing is that maximization on the simplex of  
$\sum_{1 \leq 1 \leq j \leq N} b(i,j) x_i x_j$ 
is {\bf NP-COMPLETE} even when $ b(i,j) \in \{0,1 \},1 \leq 1 \leq j \leq N$.\\
Our proof is yet another example on when the best answer to a question posed in the real numbers domain lies in the complex numbers domain.
Yet, we don't exlude a possibility of a direct ``monstrous'' proof.

\end{document}